\documentstyle[preprint,eqsecnum,aps]{revtex}
\tightenlines

\begin{document}

\begin{flushright} 

IFUP-TH 11/99 \end{flushright}
\vskip 35pt plus 3pt minus 3pt

\begin{center}
  {\large {\bf Initial State Dependence of the Breakup
of Weakly Bound Carbon Isotopes} }\footnote {\small Accepted for 
pubblication in The
Physical Review C.} 
\end{center}
\vspace{1em}
\begin{center}
{Angela Bonaccorso}\footnote 
{\small Electronic address : ANGELA.BONACCORSO@PI.INFN.IT}

{\it Istituto Nazionale di Fisica Nucleare,
 Sezione di Pisa, 56100 Pisa, Italy,}

\end{center}

\begin{center}
{\bf Abstract}
\end{center}
 {\small The one-neutron nuclear breakup from the Carbon isotopes $^{19}$C and 
$^{17}$C,  is calculated as an example of application of the theory of transfer
to the continuum reactions in the formulation which includes  spin coupling.
 The effect of the energy sharing between the parallel and transverse
neutron momentum distributions is  taken into account thus resulting in a 
theory which is more general than sudden eikonal approaches. Both effects
are necessary to understand properly the  breakup from not too weakly bound
$l_i>1$ orbitals. Breakup which leaves the core into an excited state below
particle threshold is also considered. The core-target interaction is
treated in the smooth cut-off approximation. By comparing to presently
available experimental data we show how to make some hypothesis on the
quantum numbers and occupancy of the  neutron initial state. Possible
ambiguities in the interpretation of inclusive cross sections are
discussed.  } 

\begin{flushleft} 
 {\bf PACS }
number(s):25.70.Hi, 21.10Gv,25.60Ge,25.70Mn,27.20+n
 \end{flushleft}

\begin{flushleft}  {\bf Key-words}
Breakup, momentum distribution, Carbon, d-orbitals, spin.
 \end{flushleft} 

\section{\bf INTRODUCTION}

In the last ten years since the advent of Radioactive Beams (RIBs) \cite{ta} a
new phenomenon called 'nuclear halo' \cite{hjj} has appeared in nuclear physics.
There is a halo on a nucleus as $^{11}$Be   when the last neutron
 or the last couple of neutrons, as in $^{11}$Li, are very
weakly bound and in a single particle state of low angular momentum (s
or p). Then the single particle wave function has a long tail which
extends mostly outside the potential well. Because of these
characteristics the reactions initiated by such nuclei give large
reaction cross sections and  neutron breakup cross sections. Also the
ejectile parallel momentum distributions following breakup are very
narrow, typically $40-45MeV/c$. There are also some candidates for a 
proton halo, like $^{8}B$\cite{ws,IP,ne}. But because of the Coulomb
barrier which keeps the wave function localized at the interior, there  is
still not a clear experimental evidence for this phenomenon. 

More recently another radioactive nucleus $^{19}$C has been produced but
the presence of a halo is still under discussion. There is a number of
experimental and theoretical studies of this nucleus whose results point
out to a  complex picture of its structure. Because of the presence of
d-components in the neutron wave function the reaction mechanism is rather
more complicated than for a simple  s-halo state and therefore it is more
important to be able to disentangle structure effects from effects due to
the reaction dynamics. 
 
Two sets of experiments from MSU at $E_{inc}=77$ and $88A.MeV$  have given
rather large nuclear and Coulomb breakup cross sections and narrow parallel
momentum distributions \cite{db,db1}. Consistent results were obtained from
a RIKEN experiment\cite{nak} of Coulomb breakup. A distribution similar in
shape to the MSU distribution but wider  has been measured at GSI\cite{gsi}
in a nuclear breakup experiment at $910A.MeV$.   On the other hand a GANIL
experiment based on the core-breakup reaction mechanism at
$E_{inc}=30A.MeV$ gave a narrow neutron angular distribution \cite{fm}. The
first measurement of the interaction cross sections   at GANIL \cite{sl}
did not seem to support the presence of a halo, while very recently  
rather large interaction cross sections have been measured at RIKEN
\cite{new}. One of the following sections is devoted to a brief review of
the structure calculations so far published.

In this paper we study the nuclear breakup of $^{19}$C and $^{17}$C using
the theory of transfer to the continuum \cite{abvar}-\cite{tiina}. In
\cite{ab,abb} we have shown that it  is  well adapted to describe the halo
breakup and we found that our calculations were in  good agreement   with 
experimental  breakup cross sections\cite{anne}  and  the parallel momentum
distribution widths \cite{ke}.  The transfer to the continuum formalism can
deal with  any initial binding energy and angular momentum state. It is
valid in the intermediate energy domain ($E_{inc}=10-100A.MeV$) since it
treats the relative nucleus-nucleus scattering semiclassically. The neutron
transition amplitude is however treated quantum mechanically. Therefore the
method is of intermediate complexity between the DWBA approach introduced
in \cite{traut} and simplified in \cite{shl} and the eikonal-type of
approaches used by several authors \cite{esb}-\cite{eb} to treat the
special case of halo breakup. Our approach contains several improvements
with respect to previous breakup theories in particular in so far as the
calculations of the neutron and ejectile parallel momentum distributions
are concerned. One is the introduction of spin coupling factors discussed
below. Also we treat consistently the absorption and elastic breakup of the
neutron on the target via an unitary optical model S-matrix. Since we do
not make the sudden hypothesis our formalism is more general than some 
eikonal models while  reducing to an eikonal form in the limit of zero
binding energy, as it was shown in \cite{abb}. Furthermore we introduce and
study the effect of a smooth cut-off approximation in the treatment of the
ion-ion scattering.

\section {\bf REACTION MODEL}

We do not give details of the theory here but use its main final formulas.
The theory of transfer to the continuum treats on equal footing the elastic
breakup of the neutron and its absorption from the elastic channel by the
target via an  optical model final state wave function which depends on an
unitary neutron-target S-matrix. 

 The neutron breakup probability distribution in the projectile reference
frame is \begin{equation}{dP\over dk_1} \approx {1\over
2}\Sigma_{j_f}(|1-\langle S_{j_f}\rangle |^2+1-|\langle S_{j_f}\rangle |^2)
(2j_f+1)(1+R)B_{l_f,l_i}. \label{dpde}\end{equation} 

It is obtained from a quantum mechanical transition amplitude \cite{bb}
which represents the overlap between the neutron momentum distributions in
the initial and final state when the projectile core is at a distance  $d$
from the target. The projectile-target relative motion is treated
semiclassically by using a trajectory of the center of the projectile
relative to the center of the target ${\bf s}(t)={\bf d}+{\bf v}t$ with
constant velocity v in the z-direction and impact parameter {\bf d} in the
xy-plane.   $\langle S_{j_f}\rangle$ is the energy averaged and spin
dependent optical model S-matrix which describes the neutron target
interaction. Then the first term in Eq.(\ref{dpde}), proportional to
$|1-\langle S_{j_f}\rangle |^2$ gives the neutron elastic breakup or
diffraction, while the second term  proportional to $1-|\langle
S_{j_f}\rangle |^2$ gives the neutron absorption (or stripping) by the
target. $B_{l_f,l_i}$ is an elementary transfer probability which depends
on the energies  $\varepsilon_i$ and $\varepsilon_f$, momenta $\gamma_i$
and $k_f$,  and angular momenta $l_i$ and $l_f$ of the initial and final
neutron single particle states respectively and on the incident energy per
particle, $mv^2/2$ at the distance of closest approach $d$.

\begin{equation}B_{l_f,l_i}=\left [{\hbar\over mv}\right ]{1\over
k_f}|C_i|^2 {e^{-2\eta d}\over 2\eta d}M_{l_fl_i},\label{B}\end{equation}
where $M_{l_fl_i}={1\over\sqrt{\pi}}\int_0^\infty dx e^{-x^2}
P_{l_i}(X_i+B_ix^2)P_{l_f}(X_f+B_fx^2)$ and  $X_i=1+2(k_1/\gamma_i)^2$,
 $X_f=2(k_2/k_f)^2-1$, $B_i={2\eta/ d\gamma_i}$ and $B_f={2\eta/ dk_f}$.
 $k_1=(\varepsilon_f-\varepsilon_i-{1\over 2}mv^2)/(\hbar v)$
and $k_2=(\varepsilon_f-\varepsilon_i+{1\over 2}mv^2)/(\hbar
v)$ are the z-components of the neutron momentum in the initial and final
state respectively.   
 $\eta^2=k_1^2+\gamma_i^2=k_2^2-k_f^2$ is the magnitude of the transverse
component $k_\perp=i\hbar \eta$ of the neutron  momentum in the initial and
final state. $k_\perp$ is conserved during the breakup process and it is
purely imaginary because the neutron which in the initial state has
negative energy \cite{lm} is emitted through a potential barrier.  Because
of this it holds also $k_2>k_f$. It is straightforward to see  from the
definitions of these kinematical  variables that they satisfy the neutron
energy and momentum conservation. The effect of their variation on the
reaction mechanism will be discussed in the following.

In Eq.(\ref{dpde}) R is a dynamical factor which depends on several
variables of the transfer reaction namely the Q-value and the incident
energy. In the case of nucleon transfer for a given channel specified by
$(l_f,l_i)$ this factor weighs the selectivity with respect to the four
possible transfers: 

$j_i=l_i\pm {1\over 2}\to j_f=l_f\pm {1\over 2}$

$j_i=l_i\pm {1\over 2}\to j_f=l_f\mp {1\over 2}$.

The explicit form of R is
$R=D_{j_f,j_i}F(\varepsilon_f)$
where $D_{j_f,j_i}$ is given in Table I and
$F(\varepsilon_f)=-F(k_1,l_i,\eta,\gamma_i)F(k_2,l_f,\eta,k_f)$
where $F(k,l,\eta,\gamma)={2k\eta \over \gamma^2 P_l(X)}{dP_l(X)\over dX}$.
 From Table I we see that $D_{j_f,j_i}$  has a negative sign for the
spin-flip transitions $j_i=l_i\pm{1\over 2}\to j_f=l_f\mp {1\over 2} $ and
a positive sign for the opposite situation $j_i=l_i\pm{1\over 2}\to
j_f=l_f\pm {1\over 2} $. 

Eq.(\ref{dpde})  is more general than the breakup probability discussed in
\cite{abb} because it includes spin. We use it in this paper to check the
sensitivity of the  breakup cross sections and parallel momentum
distribution widths  to changes in the initial spin of the neutron. For
example in the case of a d-state  both $1d_{3/2}$ or $1d_{5/2}$ orbits
could be occupied.  The derivation of the above equations has been given by
Hashim and Brink \cite{hb} in the case of bound-state to bound-state
transfer and was extended  by us \cite{bb3} to the final continuum case.

 Finally the  cross section \cite{ab,abb}  is given by an integration over
the core-target impact parameter \begin{eqnarray}{d \sigma_{1n}\over 
dk_1}&=&C^2S  \int_0^{\infty} d{\bf d} {dP(d)\over dk_1}P_{el}(d).\label
{tc}\end{eqnarray} The total breakup cross section is obtained by
integrating over $dk_1$. $C^2S$ is the spectroscopic factor of the neutron
single particle wave function in the initial state. The factor
$P_{el}(d)=|S_{ct}|^2$ is the core survival probability in the elastic
channel written in terms of the S-matrix for the core-target scattering. 
Since the conditions for the semiclassical approximation to the relative
ion-ion scattering apply for the reactions discussed in this paper, we use
the following parameterized form which has already been discussed in
\cite{27}: 

\begin{eqnarray}
P_{el}(d)=exp(-ln2exp[(R_s-d)/a]).\label{pel}\end{eqnarray}

 When the breakup probability is not too peaked as a function of $d$, the
above form gives a better approximation to the cross section than the
strong absorption limit used in \cite{abb}. This happens if the decay
parameter $\eta$ of the breakup probability is not very small,
corresponding to a not too small initial binding,  and when the initial
angular momentum $l_i$ is different from zero. The strong absorption radius
$R_s$ \cite{ab,abb,27} is defined by $P_{el}=1/2$ and $a$ is a diffuseness
parameter whose values will be discussed in the following. 

 Eq.(\ref{tc}) gives the final neutron parallel
momentum distribution which is related by momentum conservation to the measured
ejectile momentum distribution\cite{abb}.

\subsection{Relation to sudden approaches.}

We discuss now in more detail the relation between our model and  sudden
eikonal approaches. In this paper our main interest is to clarify the
effect of a time dependent approach on the shapes and widths of the neutron
and of the ejectile parallel momentum distributions following one-neutron
breakup. The range of validity of the sudden approximation in such a
context has recently been discussed in \cite{ga}. The discussion and the
results presented there suggest that it is best suited for incident
energies larger than 150A.MeV, for very week neutron binding  and low
initial angular momentum states ($l_i=0,1$). Under the original Glauber
terminology the same approximation is often called adiabatic because the
internal relative motion of the particles is considered slow with respect
to the relative motion of the colliding nuclei. In this sense it has been
used in \cite{yab,cr,TK} where   it was found appropriate to reproduce
several other measured quantities like the ejectile angular distributions
following neutron breakup and the absolute cross sections at relativistic
energies.

 Under the sudden (or adiabatic) hypothesis the parallel momentum
distribution of the neutron in the projectile is frozen during the reaction
and its shape should be reflected by  the final measured distribution. The
available neutron final energy is all converted into parallel momentum. 
Interference effects with the transverse distribution are in this way
neglected. The sudden hypothesis means also that the whole  momentum
distribution in the initial state is sampled during the reaction, while in
our approach the kinematical conditions, expressed in the definition of
$k_1$ and $k_2$, make a selection on the part of the initial distribution
which can be sampled by the reaction. Also, in the present approach  in
order to realize the best energy matching conditions for each possible
final energy $\varepsilon_f$ of the neutron, the reaction mechanism shares
the total momentum $k_f$  between the transverse momentum component $\eta$
and the parallel component $k_2$, thus allowing the interference (cf. also
Eq.(3) of \cite{ab}) between the two corresponding distributions. The
factor $M_{l_fl_i}$ in Eq.(\ref{B}) shows explicitly how the interference
comes about. As a  consequence the measured parallel momentum distribution
might look deformed as compared to the original parallel momentum
distribution of the neutron in the initial state of the projectile.

In order to clarify the importance of the energy sharing between the
parallel and transverse components of the neutron final momentum, we show
in Figs.(1a) and (1b), corresponding to an incident energy of 20A.MeV and
88A.MeV, respectively, the following kinematical variables as functions of
$k_1$ the neutron initial momentum with respect to the projectile :  $k_2$,
the neutron final parallel  momentum component with respect to the target,
by the dotted line;  $k_f$, the magnitude of the total neutron momentum
corresponding to each neutron final continuum energy $\varepsilon_f$, by
the dashed line; $\eta$ the neutron transverse momentum component, by the
solid line. The minimum values of $k_1$ correspond to $\varepsilon_f=0MeV$.
Clearly values of all parameters corresponding to  values of
$\varepsilon_f<0$ are not accessible by breakup reactions but they would
rather correspond to transfer to a final bound state. In both figures there
is a region corresponding to very small values of $\eta$ in which
$k_2\approx k_f$. This is the region of validity of the sudden eikonal
approximation. In fact in such conditions since the transverse component of
the neutron momentum $\eta$ is very small, the total momentum $k_f$   is
all converted into parallel momentum $k_2$.  In \cite{ab} we showed indeed
that the condition $k_2\approx k_f$ was necessary to obtain the eikonal
form of the breakup amplitude.  In the same figures we show  the initial s
and d distributions of the parallel neutron momentum as a function of
$k_1$. They are obtained from 

\begin{equation}|\tilde {\psi}_i ({\bf \rho},k_1)|^2 =\Sigma_{m_i}|2
C_iY_{l_i,m_i} (\hat k_1) K_{m_i}(\eta \rho)|^2\approx |C_i|^2~ {e^{-2\eta
\rho}\over 2\eta \rho}P_{l_i}(X_i),\label{psi} \end{equation}  which is the
one-dimensional Fourier transform of the asymptotic part of the initial
state wave function \cite{abb,lm} used to get Eq.(\ref{dpde})\cite{abb}. 
$C_i$ are the initial wave function asymptotic normalization constants
given in Table II. $K_{m_i}$ are modified Bessel functions of the second
kind.  The Legendre polynomial $P_{l_i}$ and its argument  $X_i$ have been
already defined.  The initial distribution depends on $\rho$ which is the
neutron distance from the projectile center in the x-y plane perpendicular
to the relative velocity axis between the two ions which is chosen as the
z-direction. The thick solid (s-distribution) and dotdashed
(d-distribution) lines are obtained at $\rho=6.5fm$ which is close to the
strong absorption radius value in the case of the $^{19}C+^{9}Be$ reaction.
In heavy-ion collisions the strong absorption radius  is energy dependent
and it decreases increasing the beam energy. For this reason we show also
by the dotted and thin-solid line the d-distributions  calculated at
$\rho=6 ~ {\rm and} ~5fm$ respectively. In this case the distributions are
wider.  This is one of the origins for a possible widening of the widths
when the incident energy increases. The initial distributions gets also
wider by increasing the absolute value of the initial binding. Therefore  
the eikonal approximation is best justified in a  range of $k_1\approx 0$
values and for initial distributions, like the $l_i=0$ one, concentrated in
such a region. Fig.(1b) shows that such a range increases by increasing the
incident energy. On the other hand Fig.(1a) shows that at incident energies
around 20A.MeV an important part of the  initial neutron momentum
distribution corresponding to  $k_1 $ values from $-\infty$ to about $-0.5
fm^{-1}$ would not be kinematically allowed. Thus using the frozen limit
would give too wide momentum distributions and too large breakup cross
sections. This is consistent with the recent discussion in \cite{eb}. 

Figs.(2a) and (2b) show the neutron parallel distribution after breakup
from a d-orbital with $\varepsilon_i=-1.86MeV$, in the projectile reference
frame, for the two initial beam energies used in Figs.(1a) and (1b). Such
distributions are calculated in the spin-independent approach. The solid
line is the total breakup distribution obtained from the sum of the elastic
breakup (dotted line) and absorption (dotdashed line). In both cases the
distributions are deformed with respect to the initial symmetrical one. In
particular it is interesting to see in the case of $E_{inc}=20MeV$ that
elastic breakup dominates at small initial $k_1$ while absorption of the
neutron on the target is responsible for the long tail at large $k_1$ in
both cases. The total widths are very different. At $E_{inc}=20MeV$ we get
$\hbar\Delta k_1=142MeV/c$ while at $E_{inc}=88A.MeV$ we get $\hbar\Delta
k_1=177MeV/c$, also the deformation effects are less evident in the latter
case. The strong asymmetry of the distributions can therefore be seen as a
consequence of the behavior of $\eta$ as a function of $k_1$ shown in
Fig.(1a). This shows that the beam energy dependence of the widths is due
in part to  the different range of kinematically accessible variables
$k_1$, $k_2$ and $\eta$. 

\subsection{Spin effects}

  To understand now the sensitivity of the calculated spectra to the
initial state spin we show in Fig. (3a) the neutron parallel distribution
after  breakup  from a $d_{3/2}$ orbital at $E_{inc}=88A.MeV$. In Fig. (3b)
we show the distribution after breakup from an initial $d_{5/2}$ instead.
In both figures the dotted line is the elastic breakup, from the first term
of Eq.(\ref{dpde}), the dashed line is the  absorption (or stripping) from
the second term of  Eq.(\ref{dpde}). The solid line is the sum of the two
giving the  inclusive spectrum.  It is interesting to notice that  the
absorption spectrum is very similar in the two cases. This is due to the
fact that at  high energies  the absorption depends mainly on the imaginary
part of the neutron-target optical potential while it is rather insensitive
to the spin-orbit real potential.  The elastic breakup gives instead
different spectra depending on the initial spin. As a consequence the
neutron transverse distributions should also be different, and an
experimental measure of them would help determining the total angular
momentum of the initial state, as it has been already done in \cite{chulk}.
Clearly interference and spin effects  will show up best in the data when
just one initial angular momentum state is responsible for the measured
breakup. 

 In both cases the initial symmetrical $d$-distribution which has two peaks
in Fig. (1b)  has undergone a distortion because of the reaction mechanism.
The distortion is different for  the two initial states $j_i=l_i\pm{1\over
2}$. This is a quantum mechanical effect, due to the dependence of the spin
coupling factors on the reaction Q-value and then to the  final neutron
energy. It  does not have a simple classical interpretation but we can 
explain the origin of it in our formalism. In the sum over final angular
momenta $j_f$ in Eq.(\ref{dpde}), all states with $j_f=l_f+{1\over 2}$ are
favorite with respect to the  $j_f=l_f-{1\over 2}$ for each $l_f$ because
the neutron-target spin-orbit interaction is larger  the larger the angular
momentum and then the elastic scattering probability proportional to
$|1-\langle S_{j_f}\rangle|^2$ is largest. For the same reason the neutron
leaves the projectile more easily if it is in a   $j_i=l_i-{1\over 2}$
state corresponding to a smaller neutron-core spin-orbit interaction. On
the other hand  the dependence of the spin coupling factor $F_{l\to
j}={2j+1\over 2} (1+R)$ in Eq.(\ref{dpde}) on the neutron final energy is
such that the $j_f=l_f+{1\over 2}$ states are more favorite at high neutron
energy in a spin flip transition $j_i=l_i-{1\over 2}\to j_f=l_f+{1\over 2}$
while they are more favorite at lower neutron final energy in a
no-spin-flip transition like $j_i=l_i+{1\over 2}\to j_f=l_f+{1\over 2}$.
The behavior of $F_{l\to j}$ as a function of $k_1$ is shown in the two
small figures on top of Figs.(3a) and (3b) in the case of $l_f=4$. Notice
that at any $k_1$ the $F_{l\to j}$ coefficients satisfy  $F_{l\to
j_+}+F_{l\to j_{-}}={2l_f+1}$ . Such spin coupling effects, depending on
the reaction Q-value, are a generalization of those known in transfer
between bound states and discussed among others in\cite{hb,27,vo}

\section {\bf STRUCTURE OF HEAVY CARBON ISOTOPES}

The Carbon isotopes with mass number A=17-19 belong to the category of
$2s-1d$ shell nuclei whose structure is only partially understood at
present. In a simple central plus spin-orbit potential of independent
particles   the last neutron in $^{19}$C should be in a $1d_{5/2}$ state
but more accurate shell model calculations \cite{6} and relativistic
mean-field \cite{7} find that the last occupied orbit is a $2s_{1/2}$ state
with spectroscopic factor 0.58 giving a $1/2^+$ ground state. Coupled
channel calculations give a series of possibilities in which the wave
function is a s or d state coupled to the $0^+$ or $2^+$ states of the core
\cite{Ri,Ri1}. In \cite{gsi} dynamical core polarization calculations are
reported which give a $1/2^+$ ground state with 40\% occupancy for $^{18}C
(0^+)\otimes 2s_{1/2}$,  the rest of the wave function is given by the
$^{18}C(2^+)\otimes 1d_{5/2}$ configuration. In  \cite{db,db1,gsi,Ri1} 
$^{17}$C was also studied. The ground state of this nucleus could be
$0^+\otimes 1d_{3/2}$ \cite{db1,gsi}, but shell model calculation quoted in
\cite{db1} suggest also the possibility $0.16\times(2^+\otimes
2s_{1/2})+1.58\times  (2^+\otimes 1d_{5/2})$. 

In all cases  if the last neutron is in a pure single particle state, the 
possibilities $l_i=0$ or $l_i=2$ and $d_{3/2}$ or $d_{5/2}$ should be
easily distinguished by comparing theoretical calculations to the
experimental data for one neutron breakup.  However as we have mentioned
above  both states could be only partially occupied and coupled to ground
state or to core excited states. Inclusive  experimental data can contain
contributions from several core excited states  which can eventually be
discriminated by a $\gamma$-ray experiment like the one described in
\cite{prep}. Such a situation is quite common in heavy-ion induced
reactions. For example, in the case of a "normal" nucleus like $^{40}Ar$ we
showed in \cite{tiina} that there are several possible initial states
corresponding to core excited states, all contributing to the experimental
spectrum. In particular we showed that  initial states of different angular
momenta lead to  different shapes for the ejectile inclusive spectra and
that the experimental spectrum was indeed dominated by the contribution
from a $1d_{3/2}$ coupled to a core excited state. This was due to the spin
coupling effects. 

\section {\bf RESULTS}


As an application of our theory and in the attempt to shed some light on
the $^{19}$C and nearby isotope structure  we have made some sample
calculations and a preliminary comparison with presently available
experimental data. The following quantities have been investigated: 

i) one neutron nuclear breakup cross sections from
$^{19}C$ on  $^{9}Be$, $^{12}C$ and $^{208}Pb$ targets and
$^{17}C$ on $^{9}Be$ .

ii) neutron parallel momentum distributions
for the same reactions.

The initial state parameters are given in Table II. For $^{17}C$ two
initial binding energies are considered. The first is the known neutron
separation energy, the other takes into account the extra binding  due to
the first excited state at $E^*=1.77MeV$. Previous experimental and
theoretical information on $^{17}$C can be found in \cite{nol}. In the case
of  $^{19}C$ we consider four possible initial binding energies:
$\varepsilon_i=-0.24MeV$ and $-0.5MeV$ are two possible neutron binding
energies close to the values from mass  evaluation \cite{orr} and breakup
experiments \cite{db1,nak} discussed in the literature; $-1.86MeV$ and 
$-2.12MeV$ are the corresponding  binding energies of a single particle
state coupled to the  $2^+$ excited state of $^{18}C$ which has
$E^*=1.62MeV$. 

The optical potentials used to calculate the neutron-target S-matrix are 
the same used in \cite{abb}, namely Refs. \cite{pot1,pot3} for $^{9}Be,
^{12}C$ and $^{208}Pb$ respectively. For each fixed initial state the
breakup cross section absolute values are   sensitive to both the neutron
target optical potential and to the core survival probability. This effect
has been carefully analyzed in a series of
publications\cite{abb,esb,ga,jat} and it is at present being studied by
several authors including us. In particular, as it has been suggested in
\cite{esb,jat}, it is possible that it will be necessary to  modify the
parameterization of the presently available n-$^{9}Be$ optical potentials
which were fitted mainly on low or very high energy free particle cross
sections. We  estimate that this modification could reduce the cross
sections on $^{9}Be$ up to about 30\% of the values shown here, still
leaving them within the experimental uncertitude. On the other hand the
results discussed for the other targets should be unaffected since the
calculated free neutron cross sections agree well with the known
experimental data. 

 The cross section values and momentum distributions widths for the
reaction of $^{19}C$ on $^{9}Be$ at $E_{inc}=88A.MeV$  are shown in Table
III where the sharp cut off approximation to Eq.(\ref{tc}) was used with
$R_s=1.4 (A_P^{1/3}+A_T^{1/3})fm$. In Table IV we give the values obtained
by the smooth cut-off approximation Eq.(\ref{pel}) with $a=0.6fm$. All
values in the tables are obtained by setting the initial state 
spectroscopic factor $C^2S$ equal to one. Separate contributions from
elastic breakup and absorption are also given. Our sample calculations have
shown a smooth variation of the breakup cross sections with $a$, a further
increase of its value up to $0.7fm$ gives a negligible increase in the
cross sections of about 1\%. Therefore the variation in the values of
Tables III and IV gives an estimate of the possible incertitude in the
treatment of the core-target interaction.   It appears that an increase of
50\% in the absolute value of the initial binding gives  a decrease in the
breakup cross section of 50-60\% while the widths  increase by less than
30\%. The differences between the results in the case of an initial
$d_{5/2}$ or $d_{3/2}$, both taken at the same binding energy, are instead
of the order of 10\%. The effect of the smooth cut-off is negligible in the
case of an s-state with very small binding. This is because the
free-particle limit to halo breakup discussed in \cite{abb,hans} applies in
this case. In the other cases one sees that the importance of the smooth
cut-off increases with the binding energy. To give an idea of the
sensitivity of the breakup cross section on the strong absorption radius we
have varied $R_s$ to the values $R_s=7fm, ~7.5fm, ~8fm ~and~ 8.5fm$
obtaining for the cross section the following values respectively 
$\sigma_{1n}=270mb,~214mb,~184mb,~153 mb$, for an initial s-state with
binding $-0.5MeV$, and smooth cut-off with $a=0.6$ .

 To complete the discussion on $^{19}C$ we have calculated the nuclear
diffraction component of the breakup, due to the first term of
Eq.(\ref{dpde}), at $E_{inc}=67A.MeV$ for the $^{12}C$ and $^{208}Pb$
targets used in the exclusive RIKEN experiment \cite{nak}. The measured
cross sections of \cite{nak} are $82\pm 14mb$ and $1.34\pm 0.12b$
respectively. The value on the $^{12}C$ target is supposed to be due only
to the nuclear elastic breakup, while the value on the lead target is manly
due to Coulomb breakup. In \cite{nak} a spectroscopic factor of $0.67$ is
extracted for the $0^+ \otimes 2s_{1/2}$. Also the analysis in
\cite{new,TK} of the measured interaction cross section suggest a rather
large s-component. In particular the authors of \cite{new} found their
experimental results consistent with a configuration having 46\% ($
0^+\otimes 2s_{1/2}$) and 54\% ( $2^+\otimes d_{5/2}$). 

Using the option (2) for the separation energies of Table IV, which means
$0.5MeV$ for the s-state and $2.12MeV$ for the d-state,  we find a good
agreement with the RIKEN  experimental results if we assume a spectroscopic
factor of 0.6 for the s-state and of 0.4 for the d-state and sum both
contributions. Then we obtain $\sigma_{el}=93mb$ on $^{12}C$ and
$\sigma_{el}=273mb$ on the lead target. Our estimate for the Coulomb
breakup of the s-state on lead is $\sigma_{coul}=1125mb$, such that in the
latter case our total exclusive breakup cross section is
$\sigma_{tot}=1398mb$. Following the prescription \cite{ab,abb,27} 
$R_s=1.4(A_P^{1/3}+A_T^{1/3})fm$, we took $R_s=6.9fm$ for  the $^{12}C$
target and  $R_s=12fm$ for the $^{208}Pb$ target. It is interesting to
notice that we extract the same spectroscopic factor from the light and
heavy target data, thus showing that our model and the choice of parameters
used, such as $R_s$, are appropriate for the description of the nuclear
part of the breakup cross section both on a light as well as on heavy
target. 

With the same spectroscopic factors and combination of s and d states, the
results at $88A.MeV$ are given at the top of Table IV. The cross section
value is in good agreement with the recent measurements from MSU
$\sigma_{1n}=150\pm 40mb$ \cite{pri} and it is consistently smaller than
that from the relativistic energy GSI experiment $\sigma_{1n}=233\pm 51mb$
\cite{dol}. Our width is in good agreement with the MSU value
$\Gamma_{exp}=42\pm 4MeV/c$ \cite{db1} but, as expected it is smaller than
the GSI spectrum width $\Gamma_{exp}=69\pm 3MeV/c$\cite{gsi}.  Actually we
find the best agreement with the shape of the tails of the spectrum from
\cite{db1}  if we take 30\% of s-state and 70\% of d-state, in the option
(2) for the separation energies of Table III.  Clearly because of the
present incertitude on the neutron separation energy of $^{19}C$ we
conclude that the s-state can be present with  30-60\% occupation in the
$^{19}C$ ground state. 

In \cite{gsi} it has been suggested that possible discrepancies in the
measured widths from different laboratories could originate from an
incident energy dependence of the reaction mechanism.       We have
discussed in detail  such a dynamical effect in \cite{abb,bb} and in the
first part of this paper. It is however puzzling that the discrepancy in
the measured widths from $^{19}C$ breakup does not seem to be present in
the case of $^{17}$C discussed in the following. A possible explanation has
recently been proposed in \cite{mar}. 

The results for $^{17}$C breakup at $E_{inc}=84A.MeV$ are shown in Table V.
 The values in the table are again obtained with unity spectroscopic
factors and smooth cut-off. At the top of Table V are reported the values
for the cross sections and the widths of the parallel momentum distribution
obtained summing the s and d contributions, both coupled to the core $2^+$
state, which means initial separation energy of $2.50MeV$, using instead
the spectroscopic factors quoted in \cite{db1}, namely $0.16\times
(2^+\otimes 2s_{1/2})+1.58\times (2^+\otimes d_{5/2})$. Some experimental
values from \cite{db1,dol,pri} are also given. Our results are consistent
with both experiments.  From the results given in the table we see that the
breakup from an initial pure $1d_{3/2}\otimes 0^+$ state, with binding
$-0.73MeV$,  and the breakup from the state  $ 0.16\times (2^+\otimes
2s_{1/2})+1.58\times (2^+\otimes d_{5/2})$ both give reasonable agreement
with the widths of the present data although  the shape of the experimental
spectrum \cite{db1,gsi} seems to agree better with the calculation of the
latter case, when the core is excited. Our cross section $\sigma_{1n}=93mb$
is in good agreement with the recent result from MSU $\sigma_{1n}=60\pm
20mb$\cite{pri} which was obtained in an exclusive $\gamma$-ray experiment
in which the core breakup from the $2^+$ state was identified. On the other
hand it seems possible that adding to the calculated cross section for the
$ 0.16\times (2^+\otimes 2s_{1/2})+1.58\times (2^+\otimes d_{5/2})$
configuration a contribution of about 50\% from the $1d_{3/2}\otimes 0^+$ a
 better agreement with the experimental inclusive \cite{gsi} cross section
value could be  obtained.

\section {Conclusions}
 
In this paper we have applied the transfer to the continuum theory in the
formulation which includes spin to the study of the breakup of two weakly
bound Carbon isotopes for which the d-orbital is important. The present
theory can be viewed as a generalization of sudden eikonal theories which
are obtained from our formalism taking the limit of zero initial neutron
binding energy. The utility of a time-dependent  approach with spin
coupling in the treatment of breakup from d orbitals of not too weak
binding has been clarified. 

Some hypothesis on the occupancy of the s-d shells in $^{19}$C and 
$^{17}$C have been formulated by comparing  some of the existing
experimental data  with our theoretical calculations.  Our conclusion is
that in $^{19}$C the breakup neutron occupies the d state with 40-70\%
probability while the s state has a 30-60\% occupation probability. The
present incertitude on the neutron separation energy does not allow any
definite conclusion. The s state is coupled to the ground state when we get
the best agreement with the  data, therefore the total $^{19}$C spin should
be $1/2^+$. The extreme characteristics of the s-state are responsible for
the large neutron  breakup cross section and narrow ejectile parallel
momentum distribution.

 For $^{17}$C, on the basis of the available experimental spectra, the
breakup from the ($1d_{3/2} \otimes 0 ^+$) configuration seems to show up
less than the breakup of the s-d states coupled to the core $2^+$. In the
two cases the resulting spectra have similar widths, however due to the
spin coupling effects, the spectrum of the $d_{3/2}$ breakup should show a
characteristic asymmetry which does not seem to correspond to the MSU data
nor to the GSI spectrum. New data from MSU\cite{pri} will soon be available
which hopefully will help clarifying the situation. 

 Our model takes into account the fact that breakup reactions are sensitive
only to the outermost tails of the single particle initial state wave
functions which we take as Hankel functions. It would be very important to
check with more refined structure calculations whether our hypothesis on
the occupation of the s and d states are correct. It would be also very
useful to make other experiments, like the $\gamma$-rays ones of
\cite{prep,pri} in which breakup from initial core excited states can be
measured. Finally we have suggested that a study of the neutron transverse
distribution  from  the experimental point of view, as already done in
\cite{chulk} could help resolving some puzzling cases with the help of spin
dependent reaction models like the one discussed here which contains also 
interference effects between the parallel and transverse  momentum
distributions. 

I wish to thank  P.G.Hansen and A.Mengoni for communicating  their recent
results.

{\bf Figure Captions.}
\vskip .2in
{\bf Fig.1}. Initial state momentum distributions, Eq.(\ref{psi}), as a function of
$k_1$, the neutron parallel momentum component in the initial state. For an s-state
with $\rho$=6.5fm, full curve peaked at $k_1=0$. For a d-state: dotdashed, dotted
and thin solid curves are at $\rho$=6.5fm, 6fm, 5fm, respectively.  The dotted line is
$k_2$, the neutron parallel momentum component in the final state;
the dashed line is $k_f$, the neutron total final
momentum   and the solid line is $\eta$, the neutron transverse
momentum component. (a) $ E_{inc}=20A.MeV$, (b) $ E_{inc}=88A.MeV$.

{\bf Fig.2}. The  neutron final parallel momentum distribution in the projectile
reference frame for a d-state at d=6.5fm and $\varepsilon_i=-1.86MeV$ in $^{19}C$. 
(a) $E_{inc}=20A.MeV$, (b)  $E_{inc}=88A.MeV$.

{\bf Fig.3}. The  neutron final parallel distribution in the projectile reference
frame for a d-state at $E_{inc}=88A.MeV$. 
(a) $d_{3/2}$, $j_i=l_i-1/2$; (b) $d_{5/2}$, $j_i=l_i+1/2$. Top figures give the
corresponding spin coupling coefficients  $F_{l \to j}$ for $l_f=4$. Solid line
$j_i\to j_f=l_f+1/2$, dashed line 
 $j_i\to j_f=l_f-1/2$.
\newpage

\setlength{\oddsidemargin}{-1cm}
 \setlength{\textwidth}{18cm} 
\begin {center} {\bf Table  I}. Coefficients $ D_{j_f,j_i}$ 

\begin{tabular}[bht]{|ccc|}  \hline\hline
$~~~~~~j_i$&$l_i-{1\over 2}$&$l_i+{1\over 2}$ \\
$j_f~~~~~$&&\\
\hline   
$l_f-{1\over 2}$&${1\over l_il_f}$&${-1\over l_f(l_i+1)}$  \\
$l_f+{1\over 2}$&${-1\over l_i(l_f+1)}$&${1\over (l_i+1)(l_f+1)}$\\ 
\hline\hline \end{tabular}
 \end {center}

\begin {center}
{\bf Table  II.}
Initial state parameters.

\setlength{\oddsidemargin}{-1cm}
 \setlength{\textwidth}{18cm} 
\begin{tabular}[bht]{|ccc|}  \hline\hline
 {  Projectile}&{  $|\varepsilon_i|$}&{  $C_i$}\\
 &{ (MeV)}&{  $(fm^{-1/2})$}\\ \hline
{$^{17}C$}&0.73&1.06~~~~~~0.110~~~~~~0.105\\
         &2.50&2.60~~~~~~0.500~~~~~~0.470\\
{$^{19}C$}&0.24&0.65~~~~~~0.038~~~~~~0.035\\
&0.50&0.89~~~~~~0.078~~~~~~0.074\\ 
&1.86&1.94~~~~~~0.336~~~~~~0.314\\
&2.12&---~~~~~~0.390~~~~~~---\\
 \hline
 {  Initial state}&&{$2s_{1/2}~~~~~~1d_{5/2}~~~~~~1d_{3/2}$}\\ \hline \hline\end{tabular}

\vskip 1cm
\setlength{\oddsidemargin}{-1cm}
 \setlength{\textwidth}{18cm} 
\end {center}

\begin{center}{{\bf Table  III.}
$^{19}C$ results at $E_{inc}=88A.MeV$

with sharp cut-off. }

\begin{tabular}[bht]{|ccc|}  \hline\hline
{  $|\varepsilon_i|$}&{ $\hbar\Delta k_1$}&{$\sigma_{1n}$} \\ 
(MeV)&(MeV/c)&(mb) \\
\hline 
0.24&28~~109~~100~~&441~~87~~72\\
0.50&37~~130~~120~~&270~~63~~55\\
1.86&65~~171~~160~~&94~~~32~~29\\
 \hline
 {  Initial state}&{$2s_{1/2}~1d_{5/2}~1d_{3/2}$}&
{$2s_{1/2}~1d_{5/2}~1d_{3/2}$}\\ \hline\hline \end{tabular}

\end{center}
\newpage
\setlength{\oddsidemargin}{-1cm}
 \setlength{\textwidth}{18cm} 

{\bf Table  IV.}
$^{19}C$ results at $E_{inc}=88A.MeV$ with smooth cut-off and $C^2S=1$. 
Using
$ 0.6\times(0^+\otimes 2s_{1/2})+0.4\times (2^+\otimes
d_{5/2})^{(2)}$, we get
$\sigma_{1n}=200mb$ and $\hbar\Delta k_1=40MeV/c$,
while  $\sigma_{1n}=150\pm 40mb^{\cite{pri}}$, $\Gamma_{exp}=42\pm
4MeV/c^{\cite{db1}}$.

\begin{center}
\begin{tabular}[bht]{|ccc|}  \hline\hline
{  $|\varepsilon_i|$}&{ $\hbar\Delta k_1$}&{$(\sigma_{el}~~
\sigma_{abs})~~\sigma_{1n}$} \\
 (MeV)&(MeV/c)& (mb) \\ \hline 
$^{(1)}$0.24&29~~141~~132&(194~248)442~~~(42~63)105~~~(34~53)87~\\
$^{(2)}$0.50&41~~157~~148&(129~173)302~~~(31~50)~81~~~(27~45)72~\\
$^{(1)}$1.86&68~~197~~177&(53~~83)136~~~~(18~33)~51~~~~(15~29)44~\\
$^{(2)}$2.12&---~~216~~~---&---~~~~~~~~~(16~31)~47~~~~~~~---\\
 \hline
 {  Initial state}&{$2s_{1/2}~1d_{5/2}~1d_{3/2}$~}&
{$2s_{1/2}~~~~~~~~~1d_{5/2}~~~~~~~~~1d_{3/2}$}\\ \hline\hline \end{tabular}

\vskip .8in

\end{center} 

 {\bf  Table V.}
 $^{17}C$ results at $E_{inc}=84A.MeV$
 with smooth cut-off and $C^2S=1$. 
Using $0.16\times (2^+\otimes 2s_{1/2})+1.58\times (2^+\otimes
d_{5/2})$ we get 
$\sigma_{1n}=96mb$ and $\hbar\Delta k_1=142MeV/c$, while $\sigma_{1n}=60\pm
20mb^{\cite{pri}}$ and $\Gamma_{exp}=145\pm 5MeV/c^{\cite{db1}}$,
$\sigma_{1n}=129\pm 22mb^{\cite{dol}}$, $\Gamma_{exp}=141\pm
6MeV/c^{\cite{gsi,dol}}$.
\begin{center}
 \setlength{\oddsidemargin}{-1cm}
 \setlength{\textwidth}{18cm} 
 \begin{tabular}[bht]{|ccc|}  \hline
\hline{  $|\varepsilon_i|$}&{ $\hbar\Delta
k_1 $}&{$\sigma_{1n}$}\\
  (MeV)&(MeV/c)&(mb)\\ \hline   
 0.73&45 152 142&238~~~ 65 ~~~58
\\
2.50&71  191 171&127~~~ 48 ~~~40
 \\
\hline
 {Initial state}&{$2s_{1/2}~1d_{5/2}~1d_{3/2}$}&{$2s_{1/2}~1d_{5/2}~1d_{3/2}$}\\
\hline\hline \end{tabular}

\end{center}

 
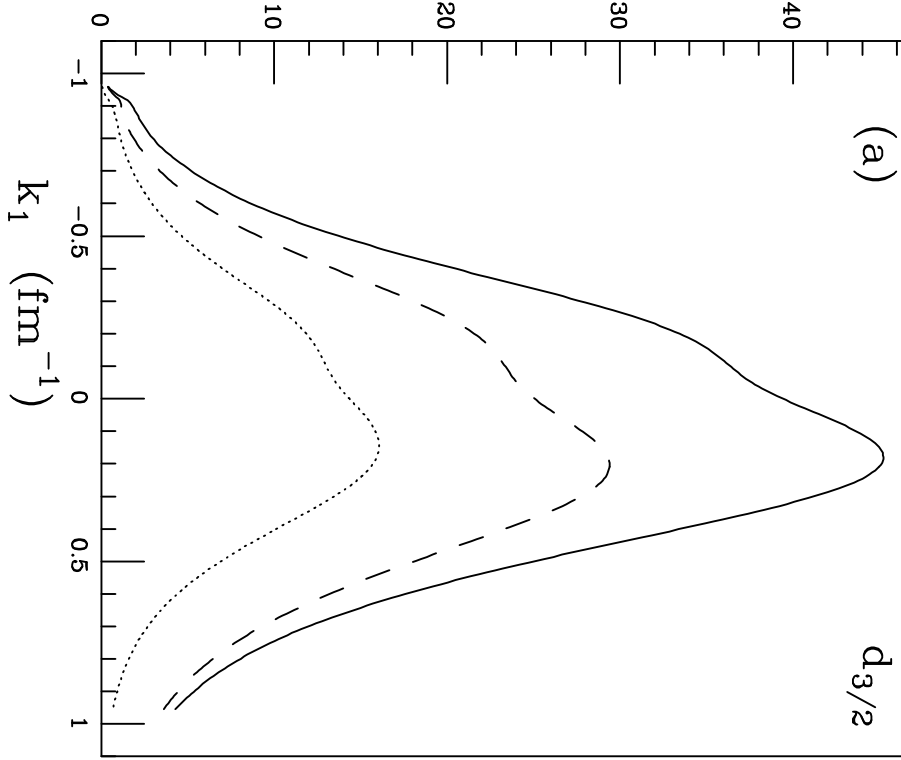
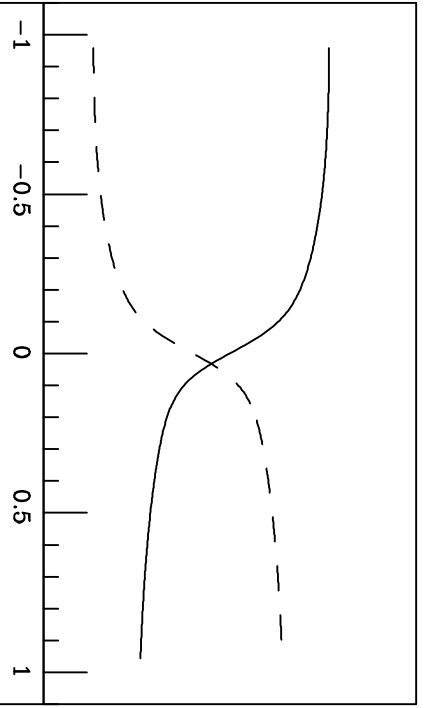
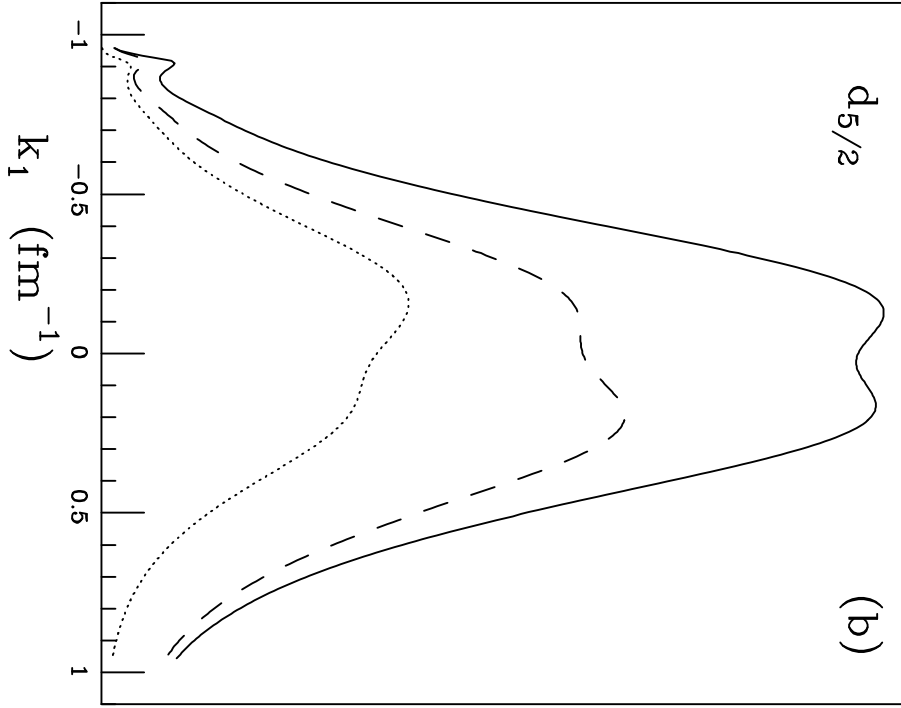
\end{document}